\documentclass[
 superscriptaddress,
 amsmath,amssymb,
aps, 
]{revtex4-2}

\usepackage{graphicx}
\usepackage{dcolumn}
\usepackage{bm}

\usepackage{xcolor}
\definecolor{ao}{rgb}{0.0, 0.5, 0.0}

\begin{document}

\preprint{APS/123-QED}

\title{
{An Alternative Scaling for} Roughness Transitions in Turbulent Flows{: The Role of the Internal Boundary Layer}
}

\author{Justin P. Cooke}
    \email{Co-Corresponding Author: justin.cooke@uri.edu}
\affiliation{%
 Mechanical Engineering and Applied Mechanics, University of Pennsylvania, Philadelphia, Pennsylvania 19104, USA
}%
\affiliation{Graduate School of Oceanography, University of Rhode Island, Kingston, Rhode Island 02881, USA }%

\author{George I. Park}
\affiliation{%
 Mechanical Engineering and Applied Mechanics, University of Pennsylvania, Philadelphia, Pennsylvania 19104, USA
}%

\author{Douglas J. Jerolmack}
\affiliation{%
 Mechanical Engineering and Applied Mechanics, University of Pennsylvania, Philadelphia, Pennsylvania 19104, USA
}%
\affiliation{%
 Department of Earth and Environmental Science, University of Pennsylvania, Philadelphia, Pennsylvania 19104, USA
}%

\author{Paulo E. Arratia}
    \email{Co-Corresponding Author: parratia@seas.upenn.edu}
\affiliation{%
 Mechanical Engineering and Applied Mechanics, University of Pennsylvania, Philadelphia, Pennsylvania 19104, USA
}%

\date{\today}

\begin{abstract} 

When turbulent boundary layer flows encounter abrupt roughness changes, an Internal Boundary Layer (IBL) forms. Equilibrium theory breaks down in the non-equilibrium IBL, which may extend O(10) km for natural atmospheric flows. Here, we find that the IBL possesses a characteristic time-scale associated with the IBL height, $\delta_i$. We show that $\delta_i$ and the edge velocity set the scales of the mean and defect velocity profiles within the IBL, for simulation and experimental data covering a multitude of roughness transition types. For the flow within the IBL, this scaling is more appropriate than more classic scaling parameters -- such as viscous and outer ones -- providing an alternative scaling for this intermediate region. We present a nontrivial extension of equilibrium theory to the dynamically adjusting IBL, which can be useful for modeling a range of environmental and industrial flows.

\end{abstract}

\maketitle


\section{Introduction}

Roughness is ubiquitous in wall-bounded turbulence~\cite{nikuradse1933stromungsgesetze,jimenez2004turbulent,chung2021predicting}. 
An Internal Boundary Layer (IBL) forms when such flows encounter abrupt changes in roughness~\cite{elliott1958growth,panofsky1964change,townsend1965response,antonia1972aresponse,antonia1972bresponse,pendergrass1984dispersion,garratt1990internal,hanson2016development,chung2021predicting,li2021experimental,gul2022experimental,cooke2024mesoscale}. At large scales, roughness transitions occur where ocean meets land, plains encounter a forest, or suburban areas transition to cities. At small scales, they can occur on ships due to patchy bio-fouling, pipes with varying internal corrosive roughness, or wind-turbine blades with leading-edge icing. At any scale, the change in wall condition results in the formation of a new flow region (\textit{i.e.}, the IBL). Within the IBL, the flow gradually adapts to the new near-wall condition~\cite{antonia1972aresponse}, whereas outside the IBL the flow retains a memory of the upstream wall condition~\cite{garratt1990internal,gul2022experimental}. The developing IBL creates non-uniform flow conditions that influence the mean velocity profile $\langle U \rangle$, turbulent kinetic energy, and wall stresses $\tau_w$~\cite{garratt1990internal}, all of which modulate near-surface transport processes including: sand and dust entrainment in deserts ~\cite{jerolmack2012internal,gunn2020macroscopic}, pollution dispersion in urban canopies ~\cite{tomas2016stable,sessa2020thermal}, and fluxes of CO$_2$, heat, vapor, and momentum~\cite{baldocchi2003assessing,savelyev2005internal,ceamanos2023remote}.

Although the flow and structure within the IBL has received much attention from the atmospheric community~\cite{garratt1990internal}, modeling flow within the IBL remains a challenge~\cite{bou2020persistent}.
Much effort has been made to capture the influence of IBL development on the mean velocity profiles. 
These have ranged from simple composite ``law of the wall'' solutions~\cite{elliott1958growth} to more complex, blended formulations~\cite{chamorro2009velocity,li2022modelling}. 
These models attempt to find a manner to describe to mean velocity profile across the boundary layer, and primarily use the IBL height to represent a vertical point of transition between functions, delineating regions inside and outside the IBL. 
Additionally, research has been devoted to characterizing and predicting the growth of the height of the IBL, $\delta_i$, which is typically described by a power law of the form $\delta_i/z_0 = A_0(\hat{x}/z_0)^{b_0}$, where $A_0$ and $b_0$ are constants, $z_0$ is a roughness parameter, and $\hat{x} = x - x_0$ is the streamwise distance from $x_0$, where the roughness transition occurs~\cite{elliott1958growth,townsend1965response,antonia1972aresponse,li2021experimental,gul2022experimental,cooke2024mesoscale}. 
The parameter $A_0$ seems to correlate with gradients in roughness; the growth exponent $b_0$ has widely varying reported values ($b_0 =$ 0.2-0.8), but is most commonly close to 0.8 ~\cite{gul2022experimental}.
{The IBL may grow larger than the boundary layer height, $\delta$, downstream, but this does not necessarily indicate flow equilibrium}~\cite{li2021experimental}.

The initial works on roughness transitions from Antonia and Luxton~\cite{antonia1972aresponse,antonia1972bresponse} indicated the flow within the IBL undergoes a phase of non-equilibrium, as the flow adjusts downstream, where it eventually reaches a state of self-preservation (equilibrium) by $\sim 20\delta_0$, where $\delta_0$ is the height of the boundary layer at the transition. 
Lee~\cite{lee2015turbulent} determined the skin-friction coefficient, $C_f$, requires a length of $\sim 30\delta_0$, which is longer than the length suggested by Anotonia and Luxton~\cite{antonia1972aresponse,antonia1972bresponse}.
Hanson and Ganapathisubramani~\cite{hanson2016development} continued to show that, due to the presence of the IBL, classical inner- and outer-scalings fail to provide an equilibrium treatment for the flow within. 
In response, they proposed a new outer-velocity scale which allows for the flow above the IBL to follow classical outer-layer similarity, and required a mixed velocity-scale to improve the near-wall scaling. 
Our understanding of the flow within the IBL still lacks the knowledge to sufficiently describe the profiles, especially before the point of equilibrium has been reached.

\begin{figure}[hbt!]
\includegraphics[width=0.75\linewidth]{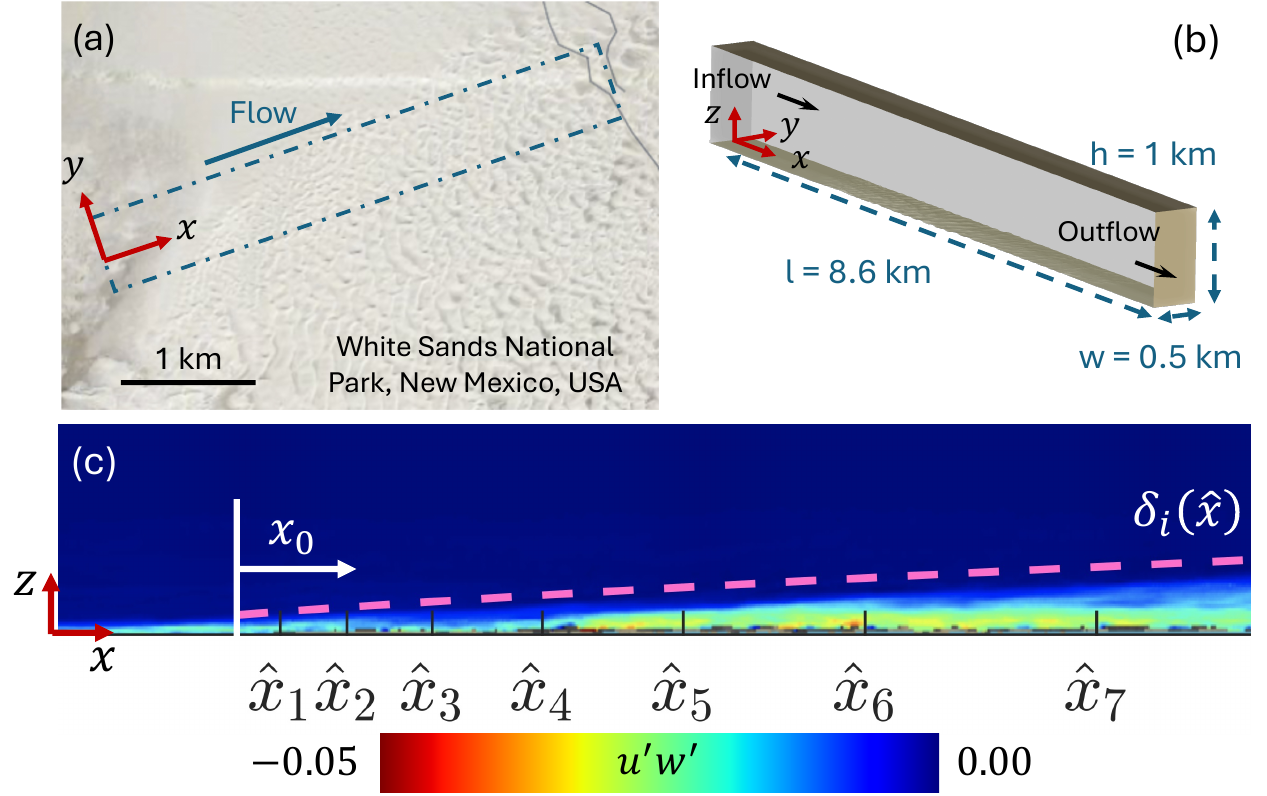}
\caption{\label{fig:setup} 
(a) Satellite imagery of White Sands National Park, New Mexico, USA. Numerical domain is outlined (blue dot-dashed box) and is aligned with the direction of dune formation. 
(b) WMLES computational domain of the simulation (S$\rightarrow$R transition) which contains the heterogeneous topography. Streamwise, spanwise, and wall-normal coordinates are in the $x$-, $y$-, and $z$-directions. (c) IBL growth shown by contour of $u'w'$ (colorbar) with IBL height $\delta_i$ (pink dashed line). 
}
\end{figure}

Recent works, which used wall-modeled large-eddy simulations (WMLES; cf., Fig.~\ref{fig:setup}b) to model an atmospheric boundary layer flow encountering a smooth to rough (S$\rightarrow$R) transition, found that turbulent stress profiles within the IBL exhibited a self-similar scaling controlled by IBL height~\cite{cooke2024mesoscale,cooke2025evolution}. 
This result suggests that there may be a common form for the mean velocity profile within the IBL, although this was not explored in these studies for a few reasons. In particular, the simulations were unable to capture the flow features in the region closest to the wall due to the high friction Reynolds number ($Re_\tau \sim O(10^6)$), limiting the observations to the log-law region and outer layer. 
In addition, it is unclear whether the self-similar behavior of the turbulent stress is limited to the sand dune flow conditions~\cite{cooke2024mesoscale,cooke2025evolution} or a general feature of IBLs. 
In this context, we attempt in the present study to seek a general scaling for the time-averaged velocity profile of turbulent flows within a developing IBL, that is applicable over different types of roughness transitions. 

In this contribution, we investigate the flow and structure of turbulent boundary layers that encounter roughness transitions in experiments and simulations. The WMLES is performed using the code \textit{CharLES}, from Cadence Design Systems (Cascade Technologies); simulation details are available in~\cite{cooke2024mesoscale,cooke2025evolution}, and the dataset is henceforth denoted ``Cooke24". We supplement the WMLES data with two experimental datasets, one from Li \textit{et al.} (denoted ``Li21'')~\cite{li2021experimental} and another from Gul and Ganapathisubramani (denoted ``Gul22'')~\cite{gul2022experimental}. 
We find the existence of a characteristic timescale associated with the IBL height and characteristic speed. 
These quantities set the scales of turbulence within, and may be used to better collapse the mean velocity profiles within the IBL immediately downstream the transition, for both simulations and experiments -- independent of the roughness transition type or other system details.
\begin{figure}[hbt!]
\includegraphics[width=0.5\linewidth]{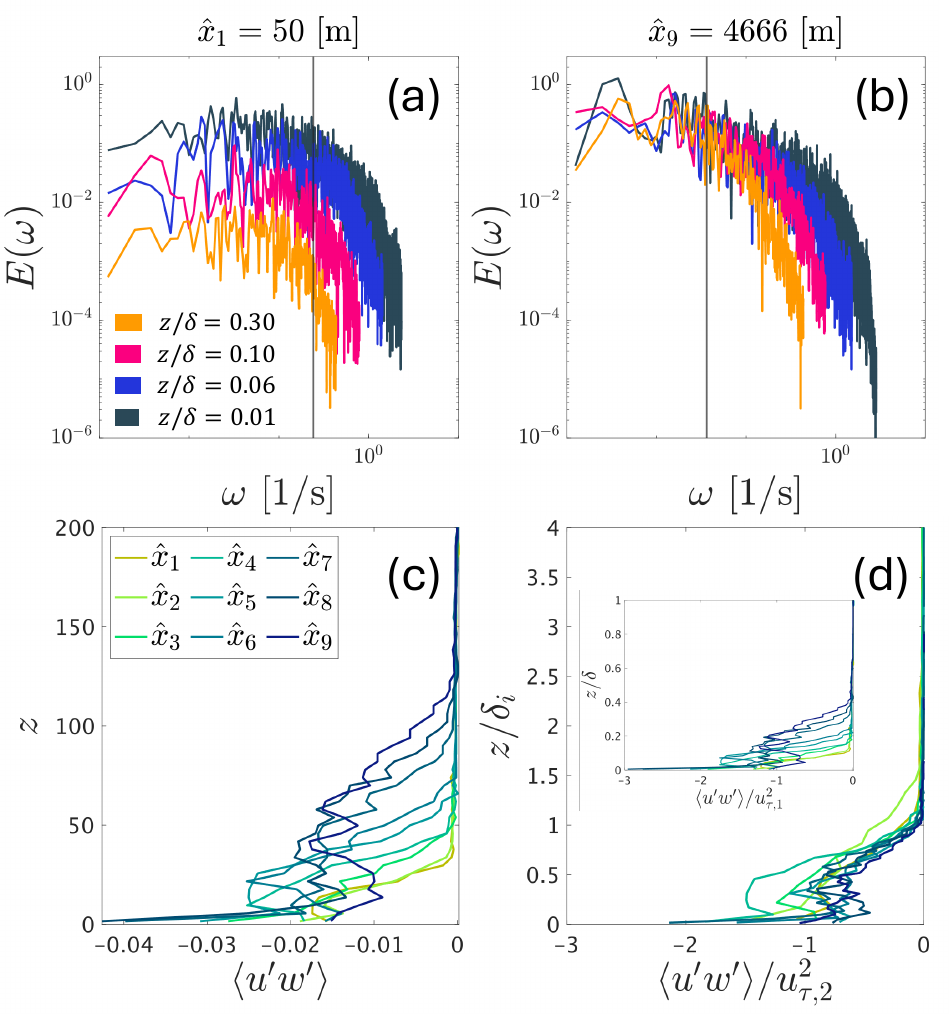}
\caption{\label{fig:prior} 
WMLES simulations from Cooke24 with a S$\rightarrow$R transition. Top: Frequency spectrum of the streamwise velocity fluctuations $E(\omega)$ captured at relative heights $z/\delta =$ 0.01 (grey), 0.06 (purple), 0.1 (pink), and 0.3 (orange) at (a) the initial station $\hat{x}_1 =$ 50 m, and (b) the final station $\hat{x}_9 =$ 4666 m. Vertical black line indicates calculated IBL frequency (vertical black line) $\omega_i = \delta_i/U_i$.  Note that low-frequency energy at the downwind station collapses, and that IBL frequency sets the scales of maximum energy. Bottom: (c) Profiles of $\langle u'w'\rangle$ downstream of the roughness transition. Legend indicates stations downwind of the transition (applied to (d) as well). (d) Same profiles as (c) normalized by the local friction velocity $u^2_{\tau,2}$ as a function of  $z/\delta_i$. For locations closest to the transition, $\hat{x}_1$ and $\hat{x}_2$, we use the atmospheric surface layer height as $\delta_i$. For all other stations farther downstream, $\delta_i$ is found using $\delta_i/z_0 = A_0(\hat{x}/z_0)^{b_0}$ with $A_0=0.29$ and $b_0=0.71$ with $z_0 = 10^{-1}$ m (see~\cite{cooke2024mesoscale} for details). 
Inset: same as (c), but normalized with $u^2_{\tau,1}$ and plot against $z/\delta$. Note that data collapses only when scaled by IBL height and $u^2_{\tau,2}$.
}
\end{figure}

\section{Details of the WMLES and Experiments}

With WMLES, we investigate the roughness transition at White Sands National Park (New Mexico, USA) where our domain captures the change from smooth salt flat to rough dune field (Fig.~\ref{fig:setup}a). This transition is expected to affect the friction velocity $u_\tau = \sqrt{\tau_w/\rho}$, where $\tau_w$ is the wall shear-stress and $\rho$ is the fluid density. Note that $u_\tau$ has different values upstream ($u_{\tau,1}$) and downstream ($u_{\tau,2}$) of the transition. The dune field topography is 
explicitly represented in the body-fitted LES calculation 
(Fig.~\ref{fig:setup}b); for more details, see~\cite{cooke2024mesoscale,cooke2025evolution}. For all datasets, the streamwise, spanwise, and wall-normal coordinates are denoted $x$, $y$, and $z$ with time-averaged velocity components $\langle U\rangle$, $\langle V\rangle$, and $\langle W\rangle$, respectively. 
Figure~\ref{fig:setup}(c) shows the growth of the instantaneous Reynolds shear-stress $u'w'$, visualizing the evolution of the IBL downstream of the roughness transition. 
The simulation also captures the growing IBL (pink dashed-line), which appears to encapsulate this turbulent region, characterizing the influence of the IBL on the flow within. 
{To quantify its growth, we follow the method of Li21~\cite{li2021experimental}, which takes streamwise differences in turbulence intensity, using the following equation:}
\begin{equation}
    \Delta \Bigg[\frac{\langle u'u'\rangle(x,z)}{U^2_\infty}\Bigg] \Big{/} \Delta \Bigg[\log_{10}\bigg(\frac{\hat{x}}{\delta}\bigg)\Bigg] \rightarrow 0.
    \label{eqn:delta_ibl}
\end{equation}

\noindent {Equation \ref{eqn:delta_ibl} describes the difference {between two locations (given by the $\Delta$ operator)} of the normalized value of $\langle u'u' \rangle$, which is a function of streamwise ($x$) and wall-normal ($z$) coordinates, divided by the normalized distance between log-spaced streamwise stations. 
The value of $\delta_i$ at the upstream streamwise station is determined as the wall-normal height in which this difference approaches zero.  
Here, $U_\infty$ is the freestream velocity.
For our simulations we choose a threshold value of $10^{-4}$ to represent convergence toward zero in Equation~\ref{eqn:delta_ibl}.
We fit a correlation of the form $\delta_i/z_0 = A_0(\hat{x}/z_0)^{b_0}$, where $A_0$ and $b_0$ are 0.29 and 0.71, respectively.}

%
\begin{figure*}[ht!]
\includegraphics[width=\linewidth]{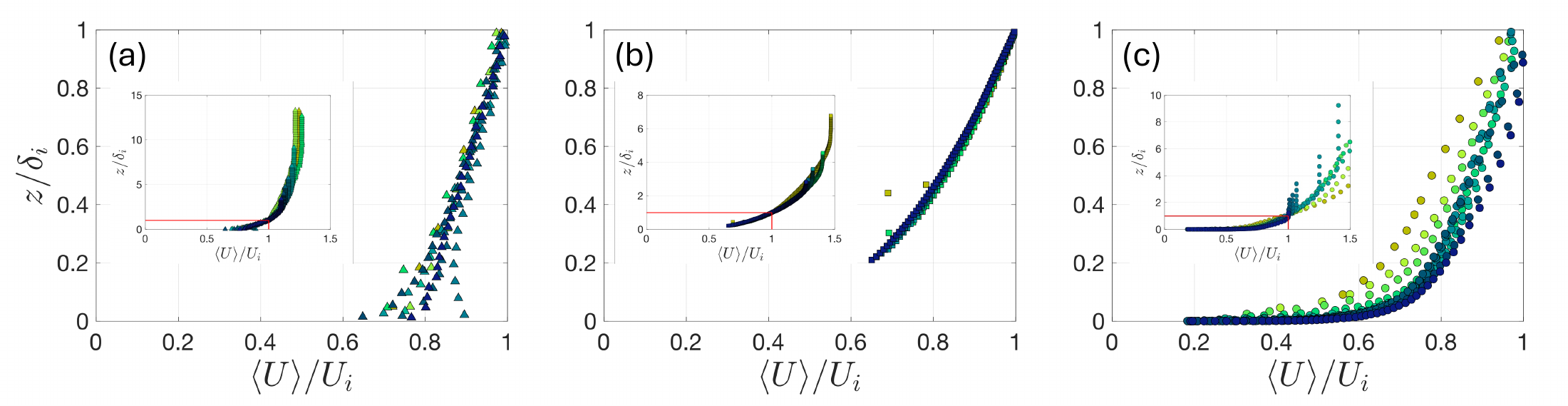}
\caption{\label{fig:meanvel} 
Velocity profiles within the IBL ($z/\delta_i <$ 1 and $\langle U \rangle/U_i <$ 1), scaled with $\delta_i$ and $U_i$. (a) Cooke24 dataset ($\triangle$) with a S$\rightarrow$R transition, (b) Gul22 dataset ($\square$) with a R$\rightarrow$Rr transition, and (c) Li21 dataset ($\circ$) with a R$\rightarrow$S transition. Inset uses the same scaling, showing the full data range of the profiles including outside the IBL. Colors are the same as Fig. \ref{fig:prior}, with yellow representing closest to $x_0$ and blue furthest.
}
\end{figure*}

Next, we compute the frequency spectrum of the streamwise velocity fluctuations $E(\omega)$ closest (Fig.~\ref{fig:prior}a) and furthest (Fig.~\ref{fig:prior}b) from the (roughness) transition at various heights ($z/\delta$). Here, $\hat{x}=x-x_0$ is the distance from the roughness transition. As expected, near $x_0$ ($\hat{x}_1=50$ m), we find that the energy decays at all frequencies as $z/\delta$ increases. However, {the energy contained within the IBL at low frequencies (for the wall-normal distances considered), remains nearly independent downstream} ($\hat{x}_9=4666$ m) {relative to the frequency spectrum found both upstream in the Alkali Flat, and shortly after the transition ($\hat{x}_1 = 50$ m).} {To confirm this, we have plot (See Supplemental Material~\cite{supp}) the frequency spectrum at each $z$, for all $\hat{x}$, and found two distinct groupings of the spectrum based on whether they fall above or below $z = \delta_i$. These low frequencies contain similar magnitudes, up to a characteristic time-scale (black line in Fig.~\ref{fig:prior}b).} Above this characteristic time-scale, the frequency spectrum follows similar behavior to the flow near $x_0$. These results indicate that the IBL possesses a characteristic frequency (or time-scale), $\omega_i$, that is associated with the internal boundary layer length-scale (height), $\delta_i$. We note that $\omega_i$ corresponds to the crossover in the energy spectrum, where energy saturates at a common magnitude for all relative heights; it also corresponds to the inverse of the expected eddy turnover time for the largest scales of turbulence coinciding to the IBL height. Thus, one can define an IBL edge velocity, $U_i = \omega_i \delta_i$.
\color{black}
The appearance of an invariant frequency (and associated scales), suggest the possibility of self-similar behavior within the IBL as it develops downstream from the roughness transition. 

To test this assumption, we first plot the Reynolds shear-stresses within the IBL, $\langle u'w'\rangle$, as a function of height and distance from $x_0$ (Fig. 2c). Results show that Reynolds shear-stress profiles thicken as the flow moves downstream from the transition point; that is, as $\hat{x}$ increases. We normalize this growth in $\langle u'w'\rangle$ using the local IBL height, $\delta_i$. This quantity is obtained using the correlation $\delta_i/z_0 = A_0(\hat{x}/z_0)^{b_0}$ with $A_0=0.29$ and $b_0=0.71$ fitted to the WMLES data, and with $z_0 = 10^{-1}$m~\cite{gunn2020macroscopic}. Figure~\ref{fig:prior}(d) shows {the normalized Reynolds shear-stresses as a function of wall-normal distance relative to IBL height. We find a suggestive collapse, particularly as the normalized Reynolds shear stresses approaches zero.} These results {suggest that the IBL set the relevant scales of the Reynolds shear-stresses by increasing their values away from the wall, and hypothesize that -- along with} the flow's longest time-scales -- the {new largest} length-scales {of the flow} are controlled by an IBL-based parameter. This implies that one can use $\delta_i$ to develop a general scaling for the velocity profile. We will employ this length-scale in a quasi-uniform approach that assumes the IBL height changes slowly enough such that the flow is in equilibrium with it, testing this hypothesis using additional experimental data. 

The Li21 dataset contains rough-to-smooth (R$\rightarrow$S) transitions for $Re_\tau$ ranging from 7000 to 21000, while the Gul22 dataset contains both R$\rightarrow$S, rough-to-rougher (R$\rightarrow$Rr), and Rr$\rightarrow$R transitions at lower $Re_\tau$ (from 1600 to 2500). From the Li21 dataset, the $Re_\tau =$ 7000 case is selected since it is a turbulent, fully rough R$\rightarrow$S transition. From the Gul22 dataset, the $Re_\tau =$ 1642 case is selected since it contains a R$\rightarrow$Rr transition. Additional experimental cases from each dataset are analyzed in the Supplemental Material~\cite{supp}, which contains Refs.~\cite{nickels2005evidence,esteban2022mean,gul2021revisiting,kundu2015fluid,materny2008experimental,smits2020some}.

\section{Alternate Scaling} 

\subsection{Mean Velocity Scaling}

We now characterize the  boundary layers from the experimental and simulation datasets by examining their mean velocity profiles, $\langle U \rangle$. Such flows can be decomposed into two regions: an inner layer and an outer layer. The inner layer ($z/\delta <$ 0.1) contains the viscous sublayer ($z^+ <$ 5), the buffer layer (5 $< z^+ <$ 50), and a portion of the log-law region ($z^+ >$ 30 and $z/\delta <$ 0.1). Here, $\cdot^+$ represents quantities scaled with viscous parameters $u_\tau$ and the fluid kinematic viscosity $\nu$. Scaling arguments from Prandtl~\cite{prandtl1925} indicate that close to the wall ($z/\delta << 1$), the mean velocity profile is a function of viscous scales, $\delta_\nu \equiv \nu/u_\tau$. Within the log-law region ($z/\delta << 1$ and $z^+ >> 1$), the dependence is lost and this function becomes constant. Conversely, the outer layer ($z^+ >$ 50) contains the rest of the log-law region, as well as a defect layer ($z/\delta >$ 0.2) where the mean velocity can become a function of $z/\delta$~\cite{coles1956law} and is controlled by outer scales ($\delta$ and $U_\infty$).
%

For smooth wall flows at equilibrium, the viscous scaling should provide a universal collapse of profiles~\cite{coles1956law}, and for rough wall flows at similar conditions, we expect an offset of the velocity profiles by some constant value (the roughness function $\Delta U^+$) in the log-layer~\cite{chung2021predicting}.
In the presence of roughness transitions and a growing IBL, these scaling methods do not adequately provide the expected collapse of profiles (see Figs. 1-8 in the Supplemental Material~\cite{supp}). Profiles very close to the wall can be collapsed using a viscous scaling, but they quickly diverge outside of the viscous sublayer; only profiles far downstream from $x_0$ display self-similarity across the boundary layer, indicating that the flow there has reached an equilibrium state. If one instead uses an outer scaling $\delta$ and $U_\infty$, we see that profiles do not immediately collapse above the defect layer ($z/\delta>0.2$); collapse only occurs closer to the boundary layer edge ($z/\delta>0.5$). These behaviors are observed in our current datasets~\cite{cooke2024mesoscale,li2021experimental,gul2022experimental} and elsewhere~\cite{antonia1972aresponse,efros2011development,hanson2016development}. A challenge is that the newly formed IBL is a nonuniform and transient flow, which inhibits the use of more classical scaling methods that rely on equilibrium assumptions. 
Thus, a new scaling that incorporates the characteristics of the IBL is necessary to adequately capture the main flow features, which in turn would allow us to understand the transport processes associated with it.    


Figure~\ref{fig:meanvel} shows the mean velocity profiles within the IBL for the three datasets as a function of normalized height and distance from $x_0$ for the WMLES S$\rightarrow$R (Fig.~\ref{fig:meanvel}a), the experimental R$\rightarrow$Rr (Fig.~\ref{fig:meanvel}b), and the experimental R$\rightarrow$S (Fig.~\ref{fig:meanvel}c) conditions. 
Here, the height and the mean velocities are normalized by $\delta_i$ and $U_i$, respectively. Remarkably, both the simulation and experimental mean velocity profiles collapse over the range $0.2 \leq z/\delta_i < 1.0$ when the IBL scales are used instead of inner or outer scaling methods. 
Note when $\delta_i < \delta$, velocity profiles diverge above $\delta_i$ (see insets of Fig.~\ref{fig:meanvel}).
Additionally, since these profiles collapse at what may be considered the defect layer of the IBL, this indicates that, for the flow within, these parameters act as an ``outer'' scale. 
{We hope to emphasize that, as with developing turbulent boundary layers,} there is still a highly non-equilibrium region close to the transition where {self-similarity may not be possible.} 
{Here, we see that the proposed} flow treatment is not as suitable {for the first three measurements stations in the Li21 dataset (}see Fig.~\ref{fig:meanvel}c{)} which contain data collected very close to $\hat{x}_0$ and within the buffer layer, which the other studies do not have. 
Nevertheless, this self-similarity is surprising considering the IBL is a highly non-equilibrium region.

\subsection{Velocity Defect Scaling}

\subsubsection{Classical Defect Scaling}

Townsend hypothesized that the outer portion of the boundary layer should resemble a wall-free turbulent flow~\cite{townsend1961equilibrium}. The idea is to use a velocity defect scaling to assess the similarity of the mean velocity profile in the outer portion of the boundary layer~\cite{connelly2006velocity}. Classically, the velocity profile over the turbulent boundary layer is taken as the sum of two functions, the ``law of the wall'' and the ``law of the wake''~\cite{coles1956law}, and is expressed in defect form as:
\begin{equation}
    \frac{U_\infty - \langle U\rangle}{u_{\tau}} = \frac{1}{\kappa}\bigg\{-ln\bigg(\frac{z}{\delta}\bigg) + \Pi\bigg[2 - w\bigg(\frac{z}{\delta}\bigg)\bigg]\bigg\},
    \label{eqn:wakelaw}
\end{equation}
where $\Pi$ is a flow dependent wake parameter, $\kappa$ is the von K\'arm\'an constant, and $w$ is the wake function given by $w(z/\delta) = 2sin^2(\pi z/2\delta)$.

\begin{figure}[bt]
\includegraphics[width=0.85\linewidth]{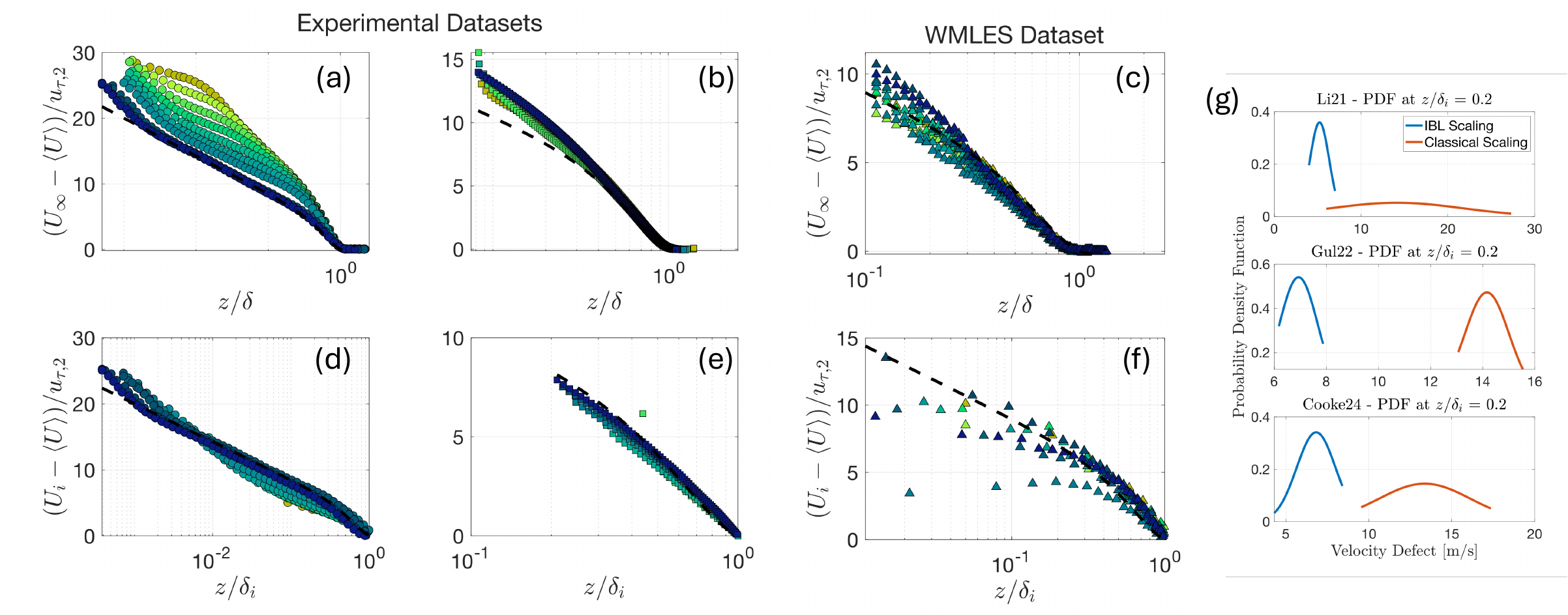}
\caption{\label{fig:defect} Velocity defect profiles using classic defect scaling (top) compared against the IBL scaling (bottom). 
The defect law (Eq.~\ref{eqn:wakelaw} black dashed line) is plotted using both $z/\delta$ and $z/\delta_i$ with $\Pi =$ 0.70 for S$\rightarrow$R and R$\rightarrow$Rr transition, and $\Pi =$ 0.55 for the R$\rightarrow$S transition.
(a,b) Li21 dataset (R$\rightarrow$S), (c,d) Gul22 dataset (R$\rightarrow$Rr), and (e,f) Cooke24 dataset (S$\rightarrow$R).
Symbols and colors are the same as Figs.~\ref{fig:prior} and \ref{fig:meanvel}.
{(g) Probability Density Functions (PDFs) (right) quantifying the collapse of the data at the beginning of the outer-layer for the IBL ($z/\delta_i \approx$ 0.2) for the classically scaled (orange lines) and IBL scaled defect profiles (blue lines).}
}
\end{figure}

Equation~\ref{eqn:wakelaw} is expected to provide a scaling that is independent of wall condition, smooth or rough~\cite{connelly2006velocity}. To test this hypothesis, we attempt to collapse velocity defect profiles according to Eq.~\ref{eqn:wakelaw} for the simulation (Fig.~\ref{fig:defect}a) and experimental datasets (Fig.~\ref{fig:defect}c,e) as a function of distance from the roughness transition ($\hat{x}$). For flows transitioning to a rougher wall condition, as in the Cooke24 (S$\rightarrow$R) and Gul22 (R$\rightarrow$Rr) datasets, we set $\Pi=0.70$ due to the stronger expected wake~\cite{kameda2008realization}, whereas for the Li21 dataset $\Pi=0.55$ is used~\cite{coles1956law}. Results show that the profiles do not collapse in the defect layer ($z/\delta > 0.2$), which is typical for outer similarity~\cite{hanson2016development}, unless the the flow is far enough from $x_0$. Similar to the mean velocity profiles, the presence of an IBL causes the conventional scaling for equilibrium boundary layer to fail. 

To address this shortcoming, we propose to introduce the IBL parameters into Eq.~\ref{eqn:wakelaw} by replacing $\delta$ and $U_\infty$ with $\delta_i$ and $U_i$. This is done at each $\hat{x}_i$ downstream from the roughness transition (see Fig.~\ref{fig:setup}c) since the IBL is growing. Despite these limitations, we find a significant improvement using the ``defect law" (Eq.~\ref{eqn:wakelaw}). {Reviewing first the Li21 dataset (Fig.~\ref{fig:defect}a,b), we find} the resulting velocity defect profiles display self-similar behavior {throughout the full inertial range of the IBL. Additionally, whereas the collapse on the defect law line was absent with classical scaling, the IBL based scaling provides an improved alignment.} 
{The profiles of the Gul22 dataset seem to initially display a reasonable collapse with classical scaling (Fig.~\ref{fig:defect}b); however, a spread between profiles is still observed, and the profiles fail to collapse onto the defect law.
As with the Li21 dataset, the profiles within the IBL are shown to agree with the defect law line when IBL-based parameters are used (Fig.~\ref{fig:defect}e).}
{Moving to the WMLES data, we note that there is} a disparity between the profiles very close to the wall (Fig.~\ref{fig:defect}c,f). We attribute this to the large scale of the roughness elements (sand dunes $\sim O(10^0-10^1)$ m in height), which creates highly locally varying flow and topography conditions that vary with position -- meaning that a profile from a specific location may not be representative of the average ~\cite{cooke2024mesoscale}. 

{To better illustrate the profile collapse within the IBL, we calculate the probability density function (PDF) of the scaled velocity data (Fig.~\ref{fig:defect}g). 
Here, the defect profiles are scaled with classical parameters (orange line) and IBL-based parameters (blue line) are taken at $z/\delta_i = 0.2$, the expected location of collapse for outer profiles in the IBL; for the classically scaled defect profiles, the equivalent $z/\delta$ height at each $\hat{x}$ is used. 
(PDFs are an effective means to quantify the probability of a random variable occurring within an interval~\cite{pope2000turbulent}.) 
Results show a tighter expected range for the defect profiles with IBL-based scaling, and this is especially evident for the Cooke24 (simulations) and Li21 (experiments) dataset. 
For the Gul22 (experiments) dataset, we detect only a slight improvement as the two scalings show similar expected ranges, with the IBL-based scaling giving a relatively smaller expected range.}
Nevertheless, all downstream profiles better follow the (IBL) defect law (Eq.~\ref{eqn:wakelaw}) {as the flow moves away from the wall and into the defect layer ($z/\delta_i > 0.2$)}, indicating the IBL parameters are a more suitable outer scale for the newly developing intermediate region. 
In summary, we have identified a new set of parameters based on the IBL that adequately collapse the velocity profiles within the developing IBL. These results can be broadly tested in other datasets since the defect law formulation remains the same. 

\section{Discussion and Conclusion} We investigated turbulent boundary layer flows undergoing roughness transitions using simulations and previously published experimental datasets. For all flows, the transition induces the formation of an Internal Boundary Layer (IBL). We find that this growing IBL prevents classical scaling methods (inner/viscous and outer) from collapsing the mean velocity and velocity defect profiles, downstream of the roughness transition ($x_0$). 
{Through examination of the frequency spectrum, we observed the existence of a characteristic frequency (time-scale) that is associated with the IBL length-scale (height), $\delta_i$. 
This time-scale is readily found experimentally and numerically. Additionally, $\delta_i$, and its corresponding velocity-scale $U_i$, were believed to be capable of providing self-similar profiles within the IBL. When tested with profiles of the Reynolds shear-stresses, our assumption was found to be true.}
Thus, a new scaling based on IBL parameters is proposed. Our results show that the local height $\delta_i$ and the edge velocity $U_i$ govern
{the outer portion of this new}
flow region and lead to self-similar behavior in both the mean velocity (Fig.~\ref{fig:meanvel}) profiles and velocity defect (Fig.~\ref{fig:defect}) profiles for all datasets. We note that this collapse occurs 
closer to the roughness transition ($x_0$) than with traditional methods~\cite{antonia1972aresponse,gul2022experimental}, suggesting that these IBL-based parameters are essential to capture the flow inside the IBL. 
Importantly, this scaling seems to be independent of roughness transition type and allows us to extend (quasi-) equilibrium methods to a non-equilibrium flow. 
These formulae can in turn be used to make informed measurements to find $U_i$, or identify it in existing datasets. 
For example, we also implemented these IBL parameters in the Zagarola-Smits scaling~\cite{zagarola1997scaling} with similarly good results (See Figs. 13-16 in the Supplemental Material~\cite{supp}). 
Our findings suggest that this newly proposed scaling can be broadly applied to roughness transition flows, and the methods described here may be useful for other types of internal boundary layers including thermal and/or pressure gradient-based. 

\begin{acknowledgements}
We are grateful for the availability of the experimental datasets by all the authors, and their care to provide the data in easily usable form. We also acknowledge Prof. Nathaniel Wei for helpful discussions which improved the manuscript. Research was supported by NASA PSTAR Award No. 80NSSC22K1313 to D.J.J. J.P.C. and G.I.P. were also funded by University of Pennsylvania Faculty Startup funds. J.P.C. acknowledges support from the Fontaine Fellowship at the University of Pennsylvania and the GEM Fellowship.
\end{acknowledgements}


\bibliography{main.bib}

\begin{thebibliography}{41}%
\makeatletter
\providecommand \@ifxundefined [1]{%
 \@ifx{#1\undefined}
}%
\providecommand \@ifnum [1]{%
 \ifnum #1\expandafter \@firstoftwo
 \else \expandafter \@secondoftwo
 \fi
}%
\providecommand \@ifx [1]{%
 \ifx #1\expandafter \@firstoftwo
 \else \expandafter \@secondoftwo
 \fi
}%
\providecommand \natexlab [1]{#1}%
\providecommand \enquote  [1]{``#1''}%
\providecommand \bibnamefont  [1]{#1}%
\providecommand \bibfnamefont [1]{#1}%
\providecommand \citenamefont [1]{#1}%
\providecommand \href@noop [0]{\@secondoftwo}%
\providecommand \href [0]{\begingroup \@sanitize@url \@href}%
\providecommand \@href[1]{\@@startlink{#1}\@@href}%
\providecommand \@@href[1]{\endgroup#1\@@endlink}%
\providecommand \@sanitize@url [0]{\catcode `\\12\catcode `\$12\catcode `\&12\catcode `\#12\catcode `\^12\catcode `\_12\catcode `\%12\relax}%
\providecommand \@@startlink[1]{}%
\providecommand \@@endlink[0]{}%
\providecommand \url  [0]{\begingroup\@sanitize@url \@url }%
\providecommand \@url [1]{\endgroup\@href {#1}{\urlprefix }}%
\providecommand \urlprefix  [0]{URL }%
\providecommand \Eprint [0]{\href }%
\providecommand \doibase [0]{https://doi.org/}%
\providecommand \selectlanguage [0]{\@gobble}%
\providecommand \bibinfo  [0]{\@secondoftwo}%
\providecommand \bibfield  [0]{\@secondoftwo}%
\providecommand \translation [1]{[#1]}%
\providecommand \BibitemOpen [0]{}%
\providecommand \bibitemStop [0]{}%
\providecommand \bibitemNoStop [0]{.\EOS\space}%
\providecommand \EOS [0]{\spacefactor3000\relax}%
\providecommand \BibitemShut  [1]{\csname bibitem#1\endcsname}%
\let\auto@bib@innerbib\@empty
\bibitem [{\citenamefont {Nikuradse}(1933)}]{nikuradse1933stromungsgesetze}%
  \BibitemOpen
  \bibfield  {author} {\bibinfo {author} {\bibfnamefont {J.}~\bibnamefont {Nikuradse}},\ }\bibfield  {title} {\bibinfo {title} {Stromungsgesetze in rauhen rohren},\ }\href@noop {} {\bibfield  {journal} {\bibinfo  {journal} {Forschung. Arb. Ing. Wes.}\ }\textbf {\bibinfo {volume} {361}},\ \bibinfo {pages} {1} (\bibinfo {year} {1933})}\BibitemShut {NoStop}%
\bibitem [{\citenamefont {Jim{\'e}nez}(2004)}]{jimenez2004turbulent}%
  \BibitemOpen
  \bibfield  {author} {\bibinfo {author} {\bibfnamefont {J.}~\bibnamefont {Jim{\'e}nez}},\ }\bibfield  {title} {\bibinfo {title} {Turbulent flows over rough walls},\ }\href@noop {} {\bibfield  {journal} {\bibinfo  {journal} {Annu. Rev. Fluid Mech.}\ }\textbf {\bibinfo {volume} {36}},\ \bibinfo {pages} {173} (\bibinfo {year} {2004})}\BibitemShut {NoStop}%
\bibitem [{\citenamefont {Chung}\ \emph {et~al.}(2021)\citenamefont {Chung}, \citenamefont {Hutchins}, \citenamefont {Schultz},\ and\ \citenamefont {Flack}}]{chung2021predicting}%
  \BibitemOpen
  \bibfield  {author} {\bibinfo {author} {\bibfnamefont {D.}~\bibnamefont {Chung}}, \bibinfo {author} {\bibfnamefont {N.}~\bibnamefont {Hutchins}}, \bibinfo {author} {\bibfnamefont {M.}~\bibnamefont {Schultz}},\ and\ \bibinfo {author} {\bibfnamefont {K.}~\bibnamefont {Flack}},\ }\bibfield  {title} {\bibinfo {title} {Predicting the drag of rough surfaces},\ }\href@noop {} {\bibfield  {journal} {\bibinfo  {journal} {Annu. Rev. Fluid Mech.}\ }\textbf {\bibinfo {volume} {53}},\ \bibinfo {pages} {439} (\bibinfo {year} {2021})}\BibitemShut {NoStop}%
\bibitem [{\citenamefont {Elliott}(1958)}]{elliott1958growth}%
  \BibitemOpen
  \bibfield  {author} {\bibinfo {author} {\bibfnamefont {W.~P.}\ \bibnamefont {Elliott}},\ }\bibfield  {title} {\bibinfo {title} {The growth of the atmospheric internal boundary layer},\ }\href@noop {} {\bibfield  {journal} {\bibinfo  {journal} {Trans. Am. Geophys. Union}\ }\textbf {\bibinfo {volume} {39}},\ \bibinfo {pages} {1048} (\bibinfo {year} {1958})}\BibitemShut {NoStop}%
\bibitem [{\citenamefont {Panofsky}\ and\ \citenamefont {Townsend}(1964)}]{panofsky1964change}%
  \BibitemOpen
  \bibfield  {author} {\bibinfo {author} {\bibfnamefont {H.~A.}\ \bibnamefont {Panofsky}}\ and\ \bibinfo {author} {\bibfnamefont {A.}~\bibnamefont {Townsend}},\ }\bibfield  {title} {\bibinfo {title} {Change of terrain roughness and the wind profile},\ }\href@noop {} {\bibfield  {journal} {\bibinfo  {journal} {Q. J. R. Meteorolog. Soc.}\ }\textbf {\bibinfo {volume} {90}},\ \bibinfo {pages} {147} (\bibinfo {year} {1964})}\BibitemShut {NoStop}%
\bibitem [{\citenamefont {Townsend}(1965)}]{townsend1965response}%
  \BibitemOpen
  \bibfield  {author} {\bibinfo {author} {\bibfnamefont {A.}~\bibnamefont {Townsend}},\ }\bibfield  {title} {\bibinfo {title} {The response of a turbulent boundary layer to abrupt changes in surface conditions},\ }\href@noop {} {\bibfield  {journal} {\bibinfo  {journal} {J. Fluid Mech.}\ }\textbf {\bibinfo {volume} {22}},\ \bibinfo {pages} {799} (\bibinfo {year} {1965})}\BibitemShut {NoStop}%
\bibitem [{\citenamefont {Antonia}\ and\ \citenamefont {Luxton}(1972{\natexlab{a}})}]{antonia1972aresponse}%
  \BibitemOpen
  \bibfield  {author} {\bibinfo {author} {\bibfnamefont {R.}~\bibnamefont {Antonia}}\ and\ \bibinfo {author} {\bibfnamefont {R.}~\bibnamefont {Luxton}},\ }\bibfield  {title} {\bibinfo {title} {The response of a turbulent boundary layer to a step change in surface roughness. {P}art 1. {Smooth to rough}},\ }\href@noop {} {\bibfield  {journal} {\bibinfo  {journal} {J. Fluid Mech.}\ }\textbf {\bibinfo {volume} {48}},\ \bibinfo {pages} {721} (\bibinfo {year} {1972}{\natexlab{a}})}\BibitemShut {NoStop}%
\bibitem [{\citenamefont {Antonia}\ and\ \citenamefont {Luxton}(1972{\natexlab{b}})}]{antonia1972bresponse}%
  \BibitemOpen
  \bibfield  {author} {\bibinfo {author} {\bibfnamefont {R.}~\bibnamefont {Antonia}}\ and\ \bibinfo {author} {\bibfnamefont {R.}~\bibnamefont {Luxton}},\ }\bibfield  {title} {\bibinfo {title} {The response of a turbulent boundary layer to a step change in surface roughness. {P}art 2. {Rough to smooth}},\ }\href@noop {} {\bibfield  {journal} {\bibinfo  {journal} {J. Fluid Mech.}\ }\textbf {\bibinfo {volume} {53}},\ \bibinfo {pages} {737} (\bibinfo {year} {1972}{\natexlab{b}})}\BibitemShut {NoStop}%
\bibitem [{\citenamefont {Pendergrass}\ and\ \citenamefont {Arya}(1984)}]{pendergrass1984dispersion}%
  \BibitemOpen
  \bibfield  {author} {\bibinfo {author} {\bibfnamefont {W.}~\bibnamefont {Pendergrass}}\ and\ \bibinfo {author} {\bibfnamefont {S.}~\bibnamefont {Arya}},\ }\bibfield  {title} {\bibinfo {title} {Dispersion in neutral boundary layer over a step change in surface roughness—i. mean flow and turbulence structure},\ }\href@noop {} {\bibfield  {journal} {\bibinfo  {journal} {Atmos. Environ.}\ }\textbf {\bibinfo {volume} {18}},\ \bibinfo {pages} {1267} (\bibinfo {year} {1984})}\BibitemShut {NoStop}%
\bibitem [{\citenamefont {Garratt}(1990)}]{garratt1990internal}%
  \BibitemOpen
  \bibfield  {author} {\bibinfo {author} {\bibfnamefont {J.}~\bibnamefont {Garratt}},\ }\bibfield  {title} {\bibinfo {title} {The internal boundary layer—a review},\ }\href@noop {} {\bibfield  {journal} {\bibinfo  {journal} {Bound.-Layer Meteorol.}\ }\textbf {\bibinfo {volume} {50}},\ \bibinfo {pages} {171} (\bibinfo {year} {1990})}\BibitemShut {NoStop}%
\bibitem [{\citenamefont {Hanson}\ and\ \citenamefont {Ganapathisubramani}(2016)}]{hanson2016development}%
  \BibitemOpen
  \bibfield  {author} {\bibinfo {author} {\bibfnamefont {R.}~\bibnamefont {Hanson}}\ and\ \bibinfo {author} {\bibfnamefont {B.}~\bibnamefont {Ganapathisubramani}},\ }\bibfield  {title} {\bibinfo {title} {Development of turbulent boundary layers past a step change in wall roughness},\ }\href@noop {} {\bibfield  {journal} {\bibinfo  {journal} {J. Fluid Mech.}\ }\textbf {\bibinfo {volume} {795}},\ \bibinfo {pages} {494} (\bibinfo {year} {2016})}\BibitemShut {NoStop}%
\bibitem [{\citenamefont {Li}\ \emph {et~al.}(2021)\citenamefont {Li}, \citenamefont {de~Silva}, \citenamefont {Chung}, \citenamefont {Pullin}, \citenamefont {Marusic},\ and\ \citenamefont {Hutchins}}]{li2021experimental}%
  \BibitemOpen
  \bibfield  {author} {\bibinfo {author} {\bibfnamefont {M.}~\bibnamefont {Li}}, \bibinfo {author} {\bibfnamefont {C.~M.}\ \bibnamefont {de~Silva}}, \bibinfo {author} {\bibfnamefont {D.}~\bibnamefont {Chung}}, \bibinfo {author} {\bibfnamefont {D.~I.}\ \bibnamefont {Pullin}}, \bibinfo {author} {\bibfnamefont {I.}~\bibnamefont {Marusic}},\ and\ \bibinfo {author} {\bibfnamefont {N.}~\bibnamefont {Hutchins}},\ }\bibfield  {title} {\bibinfo {title} {Experimental study of a turbulent boundary layer with a rough-to-smooth change in surface conditions at high {R}eynolds numbers},\ }\href@noop {} {\bibfield  {journal} {\bibinfo  {journal} {J. Fluid Mech.}\ }\textbf {\bibinfo {volume} {923}},\ \bibinfo {pages} {A18} (\bibinfo {year} {2021})}\BibitemShut {NoStop}%
\bibitem [{\citenamefont {Gul}\ and\ \citenamefont {Ganapathisubramani}(2022)}]{gul2022experimental}%
  \BibitemOpen
  \bibfield  {author} {\bibinfo {author} {\bibfnamefont {M.}~\bibnamefont {Gul}}\ and\ \bibinfo {author} {\bibfnamefont {B.}~\bibnamefont {Ganapathisubramani}},\ }\bibfield  {title} {\bibinfo {title} {Experimental observations on turbulent boundary layers subjected to a step change in surface roughness},\ }\href@noop {} {\bibfield  {journal} {\bibinfo  {journal} {J. Fluid Mech.}\ }\textbf {\bibinfo {volume} {947}},\ \bibinfo {pages} {A6} (\bibinfo {year} {2022})}\BibitemShut {NoStop}%
\bibitem [{\citenamefont {Cooke}\ \emph {et~al.}(2024)\citenamefont {Cooke}, \citenamefont {Jerolmack},\ and\ \citenamefont {Park}}]{cooke2024mesoscale}%
  \BibitemOpen
  \bibfield  {author} {\bibinfo {author} {\bibfnamefont {J.~P.}\ \bibnamefont {Cooke}}, \bibinfo {author} {\bibfnamefont {D.~J.}\ \bibnamefont {Jerolmack}},\ and\ \bibinfo {author} {\bibfnamefont {G.~I.}\ \bibnamefont {Park}},\ }\bibfield  {title} {\bibinfo {title} {Mesoscale structure of the atmospheric boundary layer across a natural roughness transition},\ }\href@noop {} {\bibfield  {journal} {\bibinfo  {journal} {PNAS}\ }\textbf {\bibinfo {volume} {121}},\ \bibinfo {pages} {2320216121} (\bibinfo {year} {2024})}\BibitemShut {NoStop}%
\bibitem [{\citenamefont {Jerolmack}\ \emph {et~al.}(2012)\citenamefont {Jerolmack}, \citenamefont {Ewing}, \citenamefont {Falcini}, \citenamefont {Martin}, \citenamefont {Masteller}, \citenamefont {Phillips}, \citenamefont {Reitz},\ and\ \citenamefont {Buynevich}}]{jerolmack2012internal}%
  \BibitemOpen
  \bibfield  {author} {\bibinfo {author} {\bibfnamefont {D.~J.}\ \bibnamefont {Jerolmack}}, \bibinfo {author} {\bibfnamefont {R.~C.}\ \bibnamefont {Ewing}}, \bibinfo {author} {\bibfnamefont {F.}~\bibnamefont {Falcini}}, \bibinfo {author} {\bibfnamefont {R.~L.}\ \bibnamefont {Martin}}, \bibinfo {author} {\bibfnamefont {C.}~\bibnamefont {Masteller}}, \bibinfo {author} {\bibfnamefont {C.}~\bibnamefont {Phillips}}, \bibinfo {author} {\bibfnamefont {M.~D.}\ \bibnamefont {Reitz}},\ and\ \bibinfo {author} {\bibfnamefont {I.}~\bibnamefont {Buynevich}},\ }\bibfield  {title} {\bibinfo {title} {Internal boundary layer model for the evolution of desert dune fields},\ }\href@noop {} {\bibfield  {journal} {\bibinfo  {journal} {Nat. Geosci.}\ }\textbf {\bibinfo {volume} {5}},\ \bibinfo {pages} {206} (\bibinfo {year} {2012})}\BibitemShut {NoStop}%
\bibitem [{\citenamefont {Gunn}\ \emph {et~al.}(2020)\citenamefont {Gunn}, \citenamefont {Schmutz}, \citenamefont {Wanker}, \citenamefont {Edmonds}, \citenamefont {Ewing},\ and\ \citenamefont {Jerolmack}}]{gunn2020macroscopic}%
  \BibitemOpen
  \bibfield  {author} {\bibinfo {author} {\bibfnamefont {A.}~\bibnamefont {Gunn}}, \bibinfo {author} {\bibfnamefont {P.}~\bibnamefont {Schmutz}}, \bibinfo {author} {\bibfnamefont {M.}~\bibnamefont {Wanker}}, \bibinfo {author} {\bibfnamefont {D.}~\bibnamefont {Edmonds}}, \bibinfo {author} {\bibfnamefont {R.}~\bibnamefont {Ewing}},\ and\ \bibinfo {author} {\bibfnamefont {D.~J.}\ \bibnamefont {Jerolmack}},\ }\bibfield  {title} {\bibinfo {title} {Macroscopic flow disequilibrium over aeolian dune fields},\ }\href@noop {} {\bibfield  {journal} {\bibinfo  {journal} {Geophys. Res. Lett.}\ }\textbf {\bibinfo {volume} {47}},\ \bibinfo {pages} {e2020GL088773} (\bibinfo {year} {2020})}\BibitemShut {NoStop}%
\bibitem [{\citenamefont {Tomas}\ \emph {et~al.}(2016)\citenamefont {Tomas}, \citenamefont {Pourquie},\ and\ \citenamefont {Jonker}}]{tomas2016stable}%
  \BibitemOpen
  \bibfield  {author} {\bibinfo {author} {\bibfnamefont {J.}~\bibnamefont {Tomas}}, \bibinfo {author} {\bibfnamefont {M.}~\bibnamefont {Pourquie}},\ and\ \bibinfo {author} {\bibfnamefont {H.}~\bibnamefont {Jonker}},\ }\bibfield  {title} {\bibinfo {title} {Stable stratification effects on flow and pollutant dispersion in boundary layers entering a generic urban environment},\ }\href@noop {} {\bibfield  {journal} {\bibinfo  {journal} {Bound.-Layer Meteorol.}\ }\textbf {\bibinfo {volume} {159}},\ \bibinfo {pages} {221} (\bibinfo {year} {2016})}\BibitemShut {NoStop}%
\bibitem [{\citenamefont {Sessa}\ \emph {et~al.}(2020)\citenamefont {Sessa}, \citenamefont {Xie},\ and\ \citenamefont {Herring}}]{sessa2020thermal}%
  \BibitemOpen
  \bibfield  {author} {\bibinfo {author} {\bibfnamefont {V.}~\bibnamefont {Sessa}}, \bibinfo {author} {\bibfnamefont {Z.-T.}\ \bibnamefont {Xie}},\ and\ \bibinfo {author} {\bibfnamefont {S.}~\bibnamefont {Herring}},\ }\bibfield  {title} {\bibinfo {title} {Thermal stratification effects on turbulence and dispersion in internal and external boundary layers},\ }\href@noop {} {\bibfield  {journal} {\bibinfo  {journal} {Bound.-Layer Meteorol.}\ }\textbf {\bibinfo {volume} {176}},\ \bibinfo {pages} {61} (\bibinfo {year} {2020})}\BibitemShut {NoStop}%
\bibitem [{\citenamefont {Baldocchi}(2003)}]{baldocchi2003assessing}%
  \BibitemOpen
  \bibfield  {author} {\bibinfo {author} {\bibfnamefont {D.~D.}\ \bibnamefont {Baldocchi}},\ }\bibfield  {title} {\bibinfo {title} {Assessing the eddy covariance technique for evaluating carbon dioxide exchange rates of ecosystems: past, present and future},\ }\href@noop {} {\bibfield  {journal} {\bibinfo  {journal} {Global Change Biol.}\ }\textbf {\bibinfo {volume} {9}},\ \bibinfo {pages} {479} (\bibinfo {year} {2003})}\BibitemShut {NoStop}%
\bibitem [{\citenamefont {Savelyev}\ and\ \citenamefont {Taylor}(2005)}]{savelyev2005internal}%
  \BibitemOpen
  \bibfield  {author} {\bibinfo {author} {\bibfnamefont {S.~A.}\ \bibnamefont {Savelyev}}\ and\ \bibinfo {author} {\bibfnamefont {P.~A.}\ \bibnamefont {Taylor}},\ }\bibfield  {title} {\bibinfo {title} {Internal boundary layers: I. height formulae for neutral and diabatic flows},\ }\href@noop {} {\bibfield  {journal} {\bibinfo  {journal} {Bound.-Layer Meteorol.}\ }\textbf {\bibinfo {volume} {115}},\ \bibinfo {pages} {1} (\bibinfo {year} {2005})}\BibitemShut {NoStop}%
\bibitem [{\citenamefont {Ceamanos}\ \emph {et~al.}(2023)\citenamefont {Ceamanos}, \citenamefont {Coopman}, \citenamefont {George}, \citenamefont {Riedi}, \citenamefont {Parrington},\ and\ \citenamefont {Clerbaux}}]{ceamanos2023remote}%
  \BibitemOpen
  \bibfield  {author} {\bibinfo {author} {\bibfnamefont {X.}~\bibnamefont {Ceamanos}}, \bibinfo {author} {\bibfnamefont {Q.}~\bibnamefont {Coopman}}, \bibinfo {author} {\bibfnamefont {M.}~\bibnamefont {George}}, \bibinfo {author} {\bibfnamefont {J.}~\bibnamefont {Riedi}}, \bibinfo {author} {\bibfnamefont {M.}~\bibnamefont {Parrington}},\ and\ \bibinfo {author} {\bibfnamefont {C.}~\bibnamefont {Clerbaux}},\ }\bibfield  {title} {\bibinfo {title} {Remote sensing and model analysis of biomass burning smoke transported across the atlantic during the 2020 western us wildfire season},\ }\href@noop {} {\bibfield  {journal} {\bibinfo  {journal} {Sci. Rep.}\ }\textbf {\bibinfo {volume} {13}},\ \bibinfo {pages} {16014} (\bibinfo {year} {2023})}\BibitemShut {NoStop}%
\bibitem [{\citenamefont {Bou-Zeid}\ \emph {et~al.}(2020)\citenamefont {Bou-Zeid}, \citenamefont {Anderson}, \citenamefont {Katul},\ and\ \citenamefont {Mahrt}}]{bou2020persistent}%
  \BibitemOpen
  \bibfield  {author} {\bibinfo {author} {\bibfnamefont {E.}~\bibnamefont {Bou-Zeid}}, \bibinfo {author} {\bibfnamefont {W.}~\bibnamefont {Anderson}}, \bibinfo {author} {\bibfnamefont {G.~G.}\ \bibnamefont {Katul}},\ and\ \bibinfo {author} {\bibfnamefont {L.}~\bibnamefont {Mahrt}},\ }\bibfield  {title} {\bibinfo {title} {The persistent challenge of surface heterogeneity in boundary-layer meteorology: a review},\ }\href@noop {} {\bibfield  {journal} {\bibinfo  {journal} {Bound.-Layer Meteorol.}\ }\textbf {\bibinfo {volume} {177}},\ \bibinfo {pages} {227} (\bibinfo {year} {2020})}\BibitemShut {NoStop}%
\bibitem [{\citenamefont {Chamorro}\ and\ \citenamefont {Port{\'e}-Agel}(2009)}]{chamorro2009velocity}%
  \BibitemOpen
  \bibfield  {author} {\bibinfo {author} {\bibfnamefont {L.~P.}\ \bibnamefont {Chamorro}}\ and\ \bibinfo {author} {\bibfnamefont {F.}~\bibnamefont {Port{\'e}-Agel}},\ }\bibfield  {title} {\bibinfo {title} {Velocity and surface shear stress distributions behind a rough-to-smooth surface transition: a simple new model},\ }\href@noop {} {\bibfield  {journal} {\bibinfo  {journal} {Bound.-layer Meteorol.}\ }\textbf {\bibinfo {volume} {130}},\ \bibinfo {pages} {29} (\bibinfo {year} {2009})}\BibitemShut {NoStop}%
\bibitem [{\citenamefont {Li}\ \emph {et~al.}(2022)\citenamefont {Li}, \citenamefont {de~Silva}, \citenamefont {Chung}, \citenamefont {Pullin}, \citenamefont {Marusic},\ and\ \citenamefont {Hutchins}}]{li2022modelling}%
  \BibitemOpen
  \bibfield  {author} {\bibinfo {author} {\bibfnamefont {M.}~\bibnamefont {Li}}, \bibinfo {author} {\bibfnamefont {C.~M.}\ \bibnamefont {de~Silva}}, \bibinfo {author} {\bibfnamefont {D.}~\bibnamefont {Chung}}, \bibinfo {author} {\bibfnamefont {D.~I.}\ \bibnamefont {Pullin}}, \bibinfo {author} {\bibfnamefont {I.}~\bibnamefont {Marusic}},\ and\ \bibinfo {author} {\bibfnamefont {N.}~\bibnamefont {Hutchins}},\ }\bibfield  {title} {\bibinfo {title} {Modelling the downstream development of a turbulent boundary layer following a step change of roughness},\ }\href@noop {} {\bibfield  {journal} {\bibinfo  {journal} {J. Fluid Mech.}\ }\textbf {\bibinfo {volume} {949}},\ \bibinfo {pages} {A7} (\bibinfo {year} {2022})}\BibitemShut {NoStop}%
\bibitem [{\citenamefont {Lee}(2015)}]{lee2015turbulent}%
  \BibitemOpen
  \bibfield  {author} {\bibinfo {author} {\bibfnamefont {J.~H.}\ \bibnamefont {Lee}},\ }\bibfield  {title} {\bibinfo {title} {Turbulent boundary layer flow with a step change from smooth to rough surface},\ }\href@noop {} {\bibfield  {journal} {\bibinfo  {journal} {International Journal of Heat and Fluid Flow}\ }\textbf {\bibinfo {volume} {54}},\ \bibinfo {pages} {39} (\bibinfo {year} {2015})}\BibitemShut {NoStop}%
\bibitem [{\citenamefont {Cooke}\ \emph {et~al.}(2025)\citenamefont {Cooke}, \citenamefont {Jerolmack},\ and\ \citenamefont {Park}}]{cooke2025evolution}%
  \BibitemOpen
  \bibfield  {author} {\bibinfo {author} {\bibfnamefont {J.~P.}\ \bibnamefont {Cooke}}, \bibinfo {author} {\bibfnamefont {D.~J.}\ \bibnamefont {Jerolmack}},\ and\ \bibinfo {author} {\bibfnamefont {G.~I.}\ \bibnamefont {Park}},\ }\bibfield  {title} {\bibinfo {title} {The evolution of turbulence producing motions in the neutral abl across a natural roughness transition},\ }\href@noop {} {\bibfield  {journal} {\bibinfo  {journal} {Journal of Geophysical Research: Atmospheres}\ }\textbf {\bibinfo {volume} {130}},\ \bibinfo {pages} {e2024JD041768} (\bibinfo {year} {2025})}\BibitemShut {NoStop}%
\bibitem [{sup()}]{supp}%
  \BibitemOpen
  \href@noop {} {}\bibinfo {note} {See Supplemental Material at URL-will-be-inserted-by-publisher for details of the experimental data, full scaling analysis of the mean velocity profiles and velocity defect profiles, and inclusion of IBL parameters in an existing scaling method.}\BibitemShut {Stop}%
\bibitem [{\citenamefont {Nickels}\ \emph {et~al.}(2005)\citenamefont {Nickels}, \citenamefont {Marusic}, \citenamefont {Hafez},\ and\ \citenamefont {Chong}}]{nickels2005evidence}%
  \BibitemOpen
  \bibfield  {author} {\bibinfo {author} {\bibfnamefont {T.}~\bibnamefont {Nickels}}, \bibinfo {author} {\bibfnamefont {I.}~\bibnamefont {Marusic}}, \bibinfo {author} {\bibfnamefont {S.}~\bibnamefont {Hafez}},\ and\ \bibinfo {author} {\bibfnamefont {M.}~\bibnamefont {Chong}},\ }\bibfield  {title} {\bibinfo {title} {Evidence of the $k_1^{-1}$ law in a high-{R}eynolds-number turbulent boundary layer},\ }\href@noop {} {\bibfield  {journal} {\bibinfo  {journal} {Phys. Rev. Lett.}\ }\textbf {\bibinfo {volume} {95}},\ \bibinfo {pages} {074501} (\bibinfo {year} {2005})}\BibitemShut {NoStop}%
\bibitem [{\citenamefont {Esteban}\ \emph {et~al.}(2022)\citenamefont {Esteban}, \citenamefont {Rodr{\'\i}guez-L{\'o}pez}, \citenamefont {Ferreira},\ and\ \citenamefont {Ganapathisubramani}}]{esteban2022mean}%
  \BibitemOpen
  \bibfield  {author} {\bibinfo {author} {\bibfnamefont {L.}~\bibnamefont {Esteban}}, \bibinfo {author} {\bibfnamefont {E.}~\bibnamefont {Rodr{\'\i}guez-L{\'o}pez}}, \bibinfo {author} {\bibfnamefont {M.}~\bibnamefont {Ferreira}},\ and\ \bibinfo {author} {\bibfnamefont {B.}~\bibnamefont {Ganapathisubramani}},\ }\bibfield  {title} {\bibinfo {title} {Mean flow of turbulent boundary layers over porous substrates},\ }\href@noop {} {\bibfield  {journal} {\bibinfo  {journal} {Phys. Rev. Fluids}\ }\textbf {\bibinfo {volume} {7}},\ \bibinfo {pages} {094603} (\bibinfo {year} {2022})}\BibitemShut {NoStop}%
\bibitem [{\citenamefont {Gul}\ and\ \citenamefont {Ganapathisubramani}(2021)}]{gul2021revisiting}%
  \BibitemOpen
  \bibfield  {author} {\bibinfo {author} {\bibfnamefont {M.}~\bibnamefont {Gul}}\ and\ \bibinfo {author} {\bibfnamefont {B.}~\bibnamefont {Ganapathisubramani}},\ }\bibfield  {title} {\bibinfo {title} {Revisiting rough-wall turbulent boundary layers over sand-grain roughness},\ }\href@noop {} {\bibfield  {journal} {\bibinfo  {journal} {J. Fluid Mech.}\ }\textbf {\bibinfo {volume} {911}},\ \bibinfo {pages} {A26} (\bibinfo {year} {2021})}\BibitemShut {NoStop}%
\bibitem [{\citenamefont {Kundu}\ \emph {et~al.}(2015)\citenamefont {Kundu}, \citenamefont {Cohen},\ and\ \citenamefont {Dowling}}]{kundu2015fluid}%
  \BibitemOpen
  \bibfield  {author} {\bibinfo {author} {\bibfnamefont {P.~K.}\ \bibnamefont {Kundu}}, \bibinfo {author} {\bibfnamefont {I.~M.}\ \bibnamefont {Cohen}},\ and\ \bibinfo {author} {\bibfnamefont {D.~R.}\ \bibnamefont {Dowling}},\ }\href@noop {} {\emph {\bibinfo {title} {Fluid mechanics}}}\ (\bibinfo  {publisher} {Academic press},\ \bibinfo {year} {2015})\BibitemShut {NoStop}%
\bibitem [{\citenamefont {Materny}\ \emph {et~al.}(2008)\citenamefont {Materny}, \citenamefont {Dr{\'o}{\.z}d{\.z}}, \citenamefont {Drobniak},\ and\ \citenamefont {Elsner}}]{materny2008experimental}%
  \BibitemOpen
  \bibfield  {author} {\bibinfo {author} {\bibfnamefont {M.}~\bibnamefont {Materny}}, \bibinfo {author} {\bibfnamefont {A.}~\bibnamefont {Dr{\'o}{\.z}d{\.z}}}, \bibinfo {author} {\bibfnamefont {S.}~\bibnamefont {Drobniak}},\ and\ \bibinfo {author} {\bibfnamefont {W.}~\bibnamefont {Elsner}},\ }\bibfield  {title} {\bibinfo {title} {Experimental analysis of turbulent boundary layer under the influence of adverse pressure gradient},\ }\href@noop {} {\bibfield  {journal} {\bibinfo  {journal} {Arch. Mech.}\ }\textbf {\bibinfo {volume} {60}},\ \bibinfo {pages} {449} (\bibinfo {year} {2008})}\BibitemShut {NoStop}%
\bibitem [{\citenamefont {Smits}(2020)}]{smits2020some}%
  \BibitemOpen
  \bibfield  {author} {\bibinfo {author} {\bibfnamefont {A.~J.}\ \bibnamefont {Smits}},\ }\bibfield  {title} {\bibinfo {title} {Some observations on reynolds number scaling in wall-bounded flows},\ }\href@noop {} {\bibfield  {journal} {\bibinfo  {journal} {Phys. Rev. Fluids}\ }\textbf {\bibinfo {volume} {5}},\ \bibinfo {pages} {110514} (\bibinfo {year} {2020})}\BibitemShut {NoStop}%
\bibitem [{\citenamefont {{Prandtl}}(1925)}]{prandtl1925}%
  \BibitemOpen
  \bibfield  {author} {\bibinfo {author} {\bibfnamefont {L.}~\bibnamefont {{Prandtl}}},\ }\bibfield  {title} {\bibinfo {title} {{7. Bericht {\"u}ber Untersuchungen zur ausgebildeten Turbulenz}},\ }\href@noop {} {\bibfield  {journal} {\bibinfo  {journal} {Z. Angew. Math. und Mech.}\ }\textbf {\bibinfo {volume} {5}},\ \bibinfo {pages} {136} (\bibinfo {year} {1925})}\BibitemShut {NoStop}%
\bibitem [{\citenamefont {Coles}(1956)}]{coles1956law}%
  \BibitemOpen
  \bibfield  {author} {\bibinfo {author} {\bibfnamefont {D.}~\bibnamefont {Coles}},\ }\bibfield  {title} {\bibinfo {title} {The law of the wake in the turbulent boundary layer},\ }\href@noop {} {\bibfield  {journal} {\bibinfo  {journal} {J. Fluid Mech.}\ }\textbf {\bibinfo {volume} {1}},\ \bibinfo {pages} {191} (\bibinfo {year} {1956})}\BibitemShut {NoStop}%
\bibitem [{\citenamefont {Efros}\ and\ \citenamefont {Krogstad}(2011)}]{efros2011development}%
  \BibitemOpen
  \bibfield  {author} {\bibinfo {author} {\bibfnamefont {V.}~\bibnamefont {Efros}}\ and\ \bibinfo {author} {\bibfnamefont {P.-{\AA}.}\ \bibnamefont {Krogstad}},\ }\bibfield  {title} {\bibinfo {title} {Development of a turbulent boundary layer after a step from smooth to rough surface},\ }\href@noop {} {\bibfield  {journal} {\bibinfo  {journal} {Exp. Fluids}\ }\textbf {\bibinfo {volume} {51}},\ \bibinfo {pages} {1563} (\bibinfo {year} {2011})}\BibitemShut {NoStop}%
\bibitem [{\citenamefont {Townsend}(1961)}]{townsend1961equilibrium}%
  \BibitemOpen
  \bibfield  {author} {\bibinfo {author} {\bibfnamefont {A.}~\bibnamefont {Townsend}},\ }\bibfield  {title} {\bibinfo {title} {Equilibrium layers and wall turbulence},\ }\href@noop {} {\bibfield  {journal} {\bibinfo  {journal} {J. Fluid Mech.}\ }\textbf {\bibinfo {volume} {11}},\ \bibinfo {pages} {97} (\bibinfo {year} {1961})}\BibitemShut {NoStop}%
\bibitem [{\citenamefont {Connelly}\ \emph {et~al.}(2006)\citenamefont {Connelly}, \citenamefont {Schultz},\ and\ \citenamefont {Flack}}]{connelly2006velocity}%
  \BibitemOpen
  \bibfield  {author} {\bibinfo {author} {\bibfnamefont {J.}~\bibnamefont {Connelly}}, \bibinfo {author} {\bibfnamefont {M.}~\bibnamefont {Schultz}},\ and\ \bibinfo {author} {\bibfnamefont {K.}~\bibnamefont {Flack}},\ }\bibfield  {title} {\bibinfo {title} {Velocity-defect scaling for turbulent boundary layers with a range of relative roughness},\ }\href@noop {} {\bibfield  {journal} {\bibinfo  {journal} {Exp. Fluids}\ }\textbf {\bibinfo {volume} {40}},\ \bibinfo {pages} {188} (\bibinfo {year} {2006})}\BibitemShut {NoStop}%
\bibitem [{\citenamefont {Kameda}\ \emph {et~al.}(2008)\citenamefont {Kameda}, \citenamefont {Mochizuki}, \citenamefont {Osaka},\ and\ \citenamefont {Higaki}}]{kameda2008realization}%
  \BibitemOpen
  \bibfield  {author} {\bibinfo {author} {\bibfnamefont {T.}~\bibnamefont {Kameda}}, \bibinfo {author} {\bibfnamefont {S.}~\bibnamefont {Mochizuki}}, \bibinfo {author} {\bibfnamefont {H.}~\bibnamefont {Osaka}},\ and\ \bibinfo {author} {\bibfnamefont {K.}~\bibnamefont {Higaki}},\ }\bibfield  {title} {\bibinfo {title} {Realization of the turbulent boundary layer over the rough wall satisfied the conditions of complete similarity and its mean flow quantities},\ }\href@noop {} {\bibfield  {journal} {\bibinfo  {journal} {J. Fluid Sci. Tech.}\ }\textbf {\bibinfo {volume} {3}},\ \bibinfo {pages} {31} (\bibinfo {year} {2008})}\BibitemShut {NoStop}%
\bibitem [{\citenamefont {Pope}(2000)}]{pope2000turbulent}%
  \BibitemOpen
  \bibfield  {author} {\bibinfo {author} {\bibfnamefont {S.~B.}\ \bibnamefont {Pope}},\ }\href@noop {} {\emph {\bibinfo {title} {Turbulent Flows}}}\ (\bibinfo  {publisher} {Cambridge University Press},\ \bibinfo {year} {2000})\BibitemShut {NoStop}%
\bibitem [{\citenamefont {Zagarola}\ and\ \citenamefont {Smits}(1997)}]{zagarola1997scaling}%
  \BibitemOpen
  \bibfield  {author} {\bibinfo {author} {\bibfnamefont {M.}~\bibnamefont {Zagarola}}\ and\ \bibinfo {author} {\bibfnamefont {A.}~\bibnamefont {Smits}},\ }\bibfield  {title} {\bibinfo {title} {Scaling of the mean velocity profile for turbulent pipe flow},\ }\href@noop {} {\bibfield  {journal} {\bibinfo  {journal} {Phys. Rev. Lett.}\ }\textbf {\bibinfo {volume} {78}},\ \bibinfo {pages} {239} (\bibinfo {year} {1997})}\BibitemShut {NoStop}%
\end{thebibliography}%
\nocite{nickels2005evidence}
\nocite{esteban2022mean}
\nocite{gul2021revisiting}
\nocite{kundu2015fluid}
\nocite{materny2008experimental}
\nocite{smits2020some}

\end{document}